\documentstyle[pre,aps,psfig,twocolumn]{revtex}
\begin{document}
\draft
\flushbottom
\twocolumn[
\hsize\textwidth\columnwidth\hsize\csname @twocolumnfalse\endcsname

\title{Escape from noisy intermittent repellers}
\author{Debabrata Biswas}
\address{Theoretical Physics Division, Bhabha Atomic Research Centre,
Trombay, Mumbai 400 085, India }

\maketitle

\tightenlines
\widetext
\advance\leftskip by 57pt
\advance\rightskip by 57pt

\begin{abstract}
Intermittent or marginally-stable repellers
are commonly associated with a power law decay in the survival fraction.
We show here that the presence of weak additive noise alters the
spectrum of the Perron - Frobenius operator significantly
giving rise to exponential decays even in systems that are otherwise
regular. \\

Version : February 3, 2000
\end{abstract}

]
\narrowtext
\tightenlines

\newcommand{\be}{\begin{equation}}
\newcommand{\ee}{\end{equation}}
\newcommand{\bea}{\begin{eqnarray}}
\newcommand{\eea}{\end{eqnarray}}
\newcommand{\Lop}{{\cal L}}

There exist a variety of physical situations where one deals with the
escape of trajectories from repellers. For instance,
nuclear physicists are interested in the escape of particles
along fission channels. Similarly, in
the study of transport coefficients in two-dimensional ballistic
conductors, one has to worry about the trapping time and its relationship
with the geometry of the system \cite{jalabert_baranger_stone}.
By and large, it is now accepted that
hyperbolic (chaotic) dynamics leads to an exponential decay in the
number of trapped particles while intermittency or marginal stability
results in power law decays. Thus, a rectangular billiard table
with a hole in the wall shows a power law decrease in the
{\em survival fraction} while a (hyperbolic) enclosure created
by the intersection of three discs shows an exponential
decay in the number of trapped particles. The reason for this
difference is intuitively clear from the following argument:
Consider that there are N particles distributed uniformly in
(Birkhoff) phase space \cite{birkhoff} and there is a hole along the
wall at $q_0$ of extent $\Delta q_0$. The (average) fraction of particles
that escape at each bounce is identical for a chaotic system
and proportional to  $\Delta q_0$ since the particles remain
uniformly distributed with time. Thus the survival fraction
decays exponentially. In a marginally stable system however, an
initial uniform distribution does not remain uniform at each bounce
since individual particles tend to stick around stable islands.
A heuristic derivation of the power law decay can be found in
\cite{bauer_bertsch} and we merely remark here that the decay exponent
is often difficult to determine analytically and an interesting
advancement in this direction has been achieved recently
by Dalhqvist \cite{dahl}.

We are interested here in a situation where marginal stability or
intermittency is accompanied by weak additive noise. Such a situation
can arise for instance in an imperfectly fabricated ballistic
conductor in the shape of a triangle or stadium where reflection is no
longer specular but has additive noise. Thus :

\bea
q_{n+1} & = & f_1(q_n,p_n)  \nonumber\\
p_{n+1} & = & f_2(q_n,p_n) + \xi_n  \label{eq:noisy_bounce}
\eea

\noindent
where ${q,p}$ are the Birkhoff coordinates \cite{birkhoff},
$f_1,f_2$ are the bounce maps and $\xi_n$ is a random variable
with $\left < \xi_n \right > = 0$ having a normalized
distribution $g(\xi)$ (normally taken to be a Gaussian with
zero mean). The question that we pose is : {\em does
one expect to find a power law decay in such a situation} ? The
answer we believe is interesting and can significantly alter the
way people look at signatures of low-dimensional chaos in
various experimental situations where noise is inevitable and
often desirable. First, however, we shall consider a 1-dimensional
intermittent map and study its spectrum in the presence of
weak noise.

A trajectory in the presence of additive noise is generated by the
iteration

\be
x_{n+1} = f(x_n) + \xi_n
\ee

\noindent
where $f(x)$ is a map, $\xi_n$ is a random variable as
described above and $x_0 \in [a,b]$. An initial density
of trajectories $\phi(x)$ evolves according to the
Perron - Frobenius equation \cite{chaos_book}:

\be
(\Lop_0 \circ \phi)(x) = \int~dy~\delta(x - f(y)) \phi(y) \label{eq:noiseless}
\ee

\noindent
in the unperturbed case. Thus, the eigenvalues and eigenfunctions
of $\Lop_0$ determine the escape rate in an open systems. More specifically,
assuming that the spectrum is discreet, an initial density
$\phi(x)$ can be expanded as

\be
\phi(x) = \sum_\alpha c_\alpha \varphi_\alpha (x)
\ee

\noindent
so that the fraction of particles that survive $n$ iterates of the map is

\bea
\Gamma (n) & = & {\int_a^b dx \left (\Lop_0^n \circ \phi \right) (x)
\over \int_a^b dx~\phi(x) } = \sum_\alpha \Lambda_\alpha^n c_\alpha
{\int_a^b dx \varphi_\alpha(x) \over \int_a^b dx~ \phi(x) } \nonumber \\
& \sim & \Lambda_0^n = {\rm e}^{-n~{\rm ln}(1/\Lambda_0)}
\hspace*{0.5cm} {\rm as} \hspace*{0.25cm} n \rightarrow \infty.
\eea

\noindent
In the above, $\{\Lambda_\alpha\}$ are the eigenvalues corresponding
to the eigenfunctions $\{\varphi_\alpha (x)\}$ and $\Lambda_0$ is
the {\em leading} eigenvalue with the largest real part. The
discreetness assumption however holds only when the dynamics is
hyperbolic. In the presence of marginally stable cycles, the
spectrum has a continuous part leading to a power law decay
of correlations (in closed systems) or survival fraction (in open
systems).

The presence of additive  noise results in
a modified kernel whose formal expression is well known \cite{lasota} :

\be
(\Lop \circ \phi )(x)  =  \int~dy~\Lop (x,y) \phi(y) \label{eq:noisy}
\ee

\noindent
where

\bea
\Lop(x,y) & = &\delta_\sigma (x - f(y)) \\
\delta_\sigma(x) & = & \int \delta(x - \xi) g(\xi)~d\xi = g(x).
\eea

\noindent
As before, if the spectrum of $\Lop$ is discreet, the decay
is exponential and the leading
eigenvalue determines the asymptotic decay rate.

Note that we have so far steered clear of spill-over effects due
to noise \cite{nicolis_bala}.
The most commonly adopted technique is the use
of {\em periodic boundary conditions} which avoids spill-over
altogether. Alternately, one can work in the infinite domain
$(-\infty , \infty)$ so that natural boundary conditions
may be employed. Yet another approach is to tailor the
noise distribution so that the probability of the dynamical
variable escaping from the interval is zero. We shall, in this
paper, have occasion to use the second and third approaches
depending on the problem and it must be noted that there are
other approaches to the spill-over problem that may be more
realistic in a given situation. Note that the spectrum and
eigenfunctions of $\Lop$ can be sensitive to the choice
of boundary conditions.

With this background, we now introduce the intermittent map
\cite{dahl,others}:

\begin{equation}
f(x)=\left\{
\begin{tabular}{lll}
$x(1 + x^2)$ & $x < 0$ \\
$x\left(1+p \; (2x)^s \right)$ &  $0 \leq x < 1/2 $\\
$2x-1$ & $ x > 1/2$
\end{tabular}  \ \ ,
\right.  \label{eq:themap}
\end{equation}

\noindent
where $s > 0$ and $p > 1$. The particle is considered to escape
if $x_{n+1} < 0$ or $x_{n+1} > 1$. The map is defined in the
infinite domain so that natural boundary conditions apply
on the density. The intermittency here is due to the fact
that $f'(0) = 1$ so that the fixed point $x = 0$ is
marginally or neutrally stable.

For an initial {\em uniform} distribution of particles in $[0,1]$,
the fraction that survives one iterate is clearly the sum of
the two intervals $I_L$ and $I_R$ for which $0 \leq f(x) \leq 1$.
Similarly, the fraction that survives two iterates is the sum
of the four intervals $I_{LL}, I_{LR}, I_{RL}, I_{RR}$ for which
$0 \leq f^2(x) \leq 1$. Generalization in this case (of binary
symbolic dynamics)
is simple : the fraction that survives $n$ iterates is the
sum of the $2^n$ intervals for each of which $0 \leq f^n(x) \leq 1$.
Each of these intervals contains a periodic point and the larger
its (in)stability, the smaller the size of the interval.
Thus \cite{chaos_book} :

\be
I_{q}^{\{n\}} = {a_q\over \Lambda_q}
\ee

\noindent
where $q$ is a symbol sequence of length $n$ consisting of  $L$ and $R$
which denotes the order in which the left and right branches
(with respect to $x = 1/2$) of the map are visited, $a_q$ is a
constant and $\Lambda_q ={d\over dx} f^{n}(x) |_{x \in I_q}$
is the stability of the periodic point.
The survival fraction can thus be expressed as \cite{dahl,chaos_book} :

\bea
\Gamma(n) &  = & \sum_{q}^{\{n\}} I_q = \sum_{q}^{\{n\}} {a_q\over \Lambda_q}
\sim \sum_p \sum_{r=1}^{\infty}
{n_p~\delta_{n,rn_p} \over \left | \Lambda_p \right |^r }
= {\cal Z}_n , \nonumber \\
{\cal Z}_n & = &
{1\over 2\pi i} \int_\gamma z^{-n}~\left ( {d\over dz}
\log~\zeta^{-1}(z) \right)~dz \label{eq:relation}
\eea

\noindent
where $\zeta^{-1}(z) = \prod_p \left ( 1 - z^{n_p}/
\left| \Lambda_p \right | \right )$ is the dynamical zeta function
and $\gamma$ is a (small) negatively oriented contour around the origin.
Dahlqvist \cite{dahl} has recently shown that in the {\em noiseless}
case, $\zeta^{-1}(z)$ has a singularity of the
type $(1 - z)^{1/s}$. It then follows \cite{dahl} from eq.~(\ref{eq:relation}) that
the survival fraction, $\Gamma(n) \sim
1/ n^{1/s}$ for an initial {\em uniform} distribution of particles.

The zeta function is also (approximately) related to the eigenvalues
of $\Lop$ through the relation :

\be
{\rm Tr}~\Lop^n = \sum_\alpha \Lambda_\alpha^n = \sum_p \sum_{r=1}^{\infty}
{n_p~\delta_{n,rn_p} \over \left | 1 - \Lambda_p \right |^r }
\simeq {\cal Z}_n
\ee

\noindent
When the zeta function is analytic, its zeroes, $\{z_k\}$ are isolated
and related to $\Lambda_\alpha$ as $\Lambda_\alpha = 1/z_k$.
On the other hand, when the system is intermittent and $\zeta^{-1}(z)$
displays a branch cut, the spectrum of  $\Lop$ no longer remains discreet.
Thus intermittency leads to an asymptotic power law decay.
For $s = 0.7, p = 1.2$, the initial decay is however exponential
and this is ascribed to a pair of complex conjugate roots \cite{dahl}.
The power law behaviour emerges only after 600 iterations of the map.

\begin{figure}[tbp]
{\hspace*{0.15cm}\psfig{figure=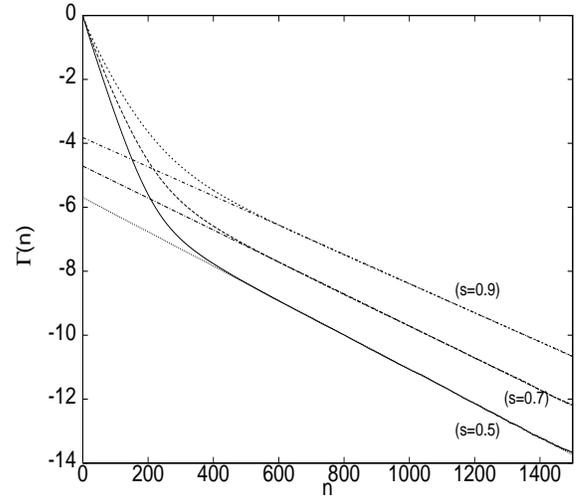,height=6.5cm,width=7.65cm,angle=270}}
{\vspace*{.13in}}
\caption[ty]{The survival fraction, $\Gamma(n)$ (log scale)
for $s = 0.5$, $0.7$ and $0.9$ with $\sigma = 0.002$.
The straight lines (dashed) are the best fit exponential decays.
The initial distribution in each case consists of $10^{10}$ points.}
\label{fig:1}
\end{figure}

We now consider the map (eq.(\ref{eq:themap})) with weak additive
Gaussian noise

\be
g(\xi) = {1\over \sqrt{2 \pi \sigma^2}} e^{-{\xi^2 \over 2\sigma^2}}
\ee

\noindent
for $s = 0.9, 0.7$ and $0.5$ and with $\sigma = 0.002$ (see figure
\ref{fig:1}).

\noindent
Clearly there is a transition from a power law
to an exponential decay in the presence of weak noise in each of
the three cases for large $n$. Note, however, that the initial decay,
though exponential, is at a significantly different rate and the
slope of $\Gamma(n)$ settles down
to the asymptotic value {\em gradually} after a large number of iterations.

Two inferences can be drawn from this transition from power-law
to exponential behaviour. First, the presence of weak
noise makes the eigenvalues (of $\Lop$, see eq.~(\ref{eq:noisy}) )
discreet. Also,  there are closely spaced eigenvalues
with small differences in their real parts around the leading
eigenvalue $\Lambda_0$ which leads to the gradual change in
slope of $\Gamma(n)$.

The discreetness of the spectrum follows from the fact
that the noisy kernel is integrable and bounded. The
corresponding Fredholm determinant \cite{integral} is thus entire
whose zeroes ($1/\Lambda_\alpha$) are isolated.
Thus, even for very weak noise, the spectrum is 
discreet though the transition time  may be too large 
for the final exponential decay to be observed 
experimentally.  

\begin{figure}[tbp]
{\hspace*{0.15cm}\psfig{figure=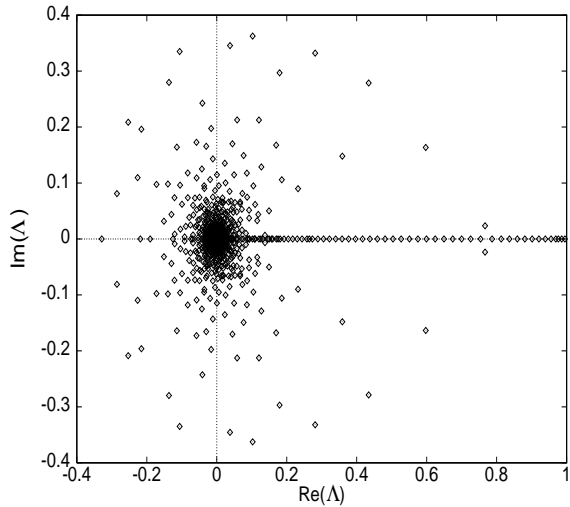,height=6.5cm,width=7.65cm,angle=270}}
{\vspace*{.13in}}
\caption[ty]{ The spectrum of $\Lop$ for $s= 0.7$ and $\sigma = 0.002$.}
\label{fig:2}
\end{figure}

The closely spaced eigenvalues around $\Lambda_0$
are possibly remnants of the continuous spectrum 
which exists for the noiseless case. In order to 
understand this better, we have evaluated the 
eigenvalues of $\Lop$ by discreetizing the
integral equation and diagonalizing the resulting 
matrix \cite{noise3}. Recall that a Fredholm
integral equation exists as a limit of discreet
sum \cite{integral} so that a matrix representation
is adequate so long as its order is large and
spurious eigenvalues are eliminated.   
Figure \ref{fig:2} shows a plot of
the eigenvalues of $\Lop$ for $s = 0.7$ and $\sigma = 0.002$
(see eq.~(\ref{eq:themap}))
obtained using a matrix of size $2500 \times 2500$.
We have checked that the relevant eigenvalues on the
positive real axis have converged to the fourth
significant digit.

\noindent
Clearly, the closely spaced eigenvalues along the real
line do not allow the survival-fraction
to be dominated by the leading eigenvalue for small $n$.
An order of magnitude evaluation of the transition time 
can be made by noting that 
$\Gamma(n) \simeq e^{-n\ln(1/\Lambda_0)} \left ( d_0 +
d_1 e^{-n \ln(\Lambda_0/\Lambda_1)} \right)$ where $\Lambda_1$ is
the next to leading eigenvalue and 
$d_\alpha = c_\alpha\int~dx \phi_\alpha(x)$.
Thus, for $n >> n_{trans} = 1/\ln(\Lambda_0/\Lambda_1)$,
the leading eigenvalue dominates. For the three values
of $s$ considered, $n_{trans} \simeq 98, 105, 119$ for
$s = 0.5, 0.7, 0.9$ respectively so that exponential decay
sets in first for $s= 0.5$ as observed in fig.~\ref{fig:1}.
Thus the difference between the leading and the next-to-leading
eigenvalue gives a good qualitative picture and fixes a lower
bound for the transition time.
Note also that for each of the three $s$-values considered, the
leading eigenvalue accurately reproduces the asymptotic
decay of the survival-fraction (see Table \ref{tab}).

We now turn our attention to the effect of noise in non-chaotic
systems. Specifically, we shall consider triangular billiards
which are non-chaotic though generically non-integrable.
The only integrable examples are the $(\pi/3,\pi/3,\pi/3)$,
$(\pi/2,\pi/3,\pi/6)$  and the $(\pi/2,\pi/4,\pi/4)$
triangles while all other rational triangles have at least one
internal angle of the form $m\pi/n, m > 1 $ and are non-integrable.
Their invariant surface, though $2$-dimensional, is not a torus but
topologically equivalent to a sphere with multiple holes. Also these
systems are non-chaotic though irrational triangles are possibly
ergodic and even display the weak mixing property \cite{artuso_hobson}.
A linear stability analysis shows that the Jacobian
matrix has unit eigenvalues and hence these billiards are
marginally stable \cite{db97}.

Consider such a triangular billiard of unit perimeter and let
${q,p}$ denote the Birkhoff co-ordinates. Here $q$ is measured
along the boundary while $p = \sin(\theta)$ where $\theta$ is
the angle between the ray and the inward normal at
the boundary point $q$. Thus $q \in [0,1]$ and $p \in [-1,1]$.
In a typical experiment, one considers an initial uniform distribution
of particles ($\sim 10^8$) in this phase space evolving freely in
between bounces and reflecting
specularly from the walls. The particles are allowed to escape
through a small opening at $q_0$ of extent $\Delta q_0 (= 0.005)$.
For both the integrable $(\pi/2,\pi/3)$ and non-integrable 
$(18\pi/31,17\pi/97)$ triangles considered, the initial
decay is exponential while the asymptotic decay is a power
law, $\Gamma(n) \sim n^{-\beta}$, with $\beta = 1.035$ in the
integrable case and $\beta = 1.085$ in the non-integrable (NI)
case. Thus pre-exponential decays are not exceptional and
can persist for a long time in marginally stable systems.

A more realistic situation should however include noise. For
instance, imperfections can give rise to maps of the
type considered in eq.~(\ref{eq:noisy_bounce}). This leads
to interesting results.
For Gaussian noise (in Birkhoff momentum)
with $\sigma = 0.000001$, a single exponential decay
dominates the survival fraction in the NI case for
$n > 5000$ while in the integrable case, the transition
continues beyond $n = 14000$. Moreover, the closely 
spaced eigenvalues leads to a quasi-algebraic decay
in the interval $6000 < n < 14000$
for the integrable case. Note that most trajectories
remain largely unaffected for several hundred bounces
for the value of $\sigma$ considered so that $\Gamma(n)$
closely follows the noiseless case initially.
Thus, even when the asymptotic decay is exponential, transition 
times can be very large.

In the weak noise case, the evolution operator 
can be approximated as 

\be
(\Lop \circ \phi) (x) \simeq \xi(x) - {\sigma^2 \over 2} \xi''(x)
\ee

\noindent
where $\xi(x) = \sum_i \phi(x)/\left | f'(f_i^{-1}(x)) \right |$.
Using a polynomial basis \cite{noise3}, a matrix representation
of the operator can be constructed where the elements
${\Lop}_{mn} = {d^m \over dx^m } \{ \Lop \circ ({x^n \over n!}) \}$.
Writing $\sigma$ as $\sigma_1 + \sigma_0$ where $\sigma_0 
\simeq 0^+$, a perturbation
calculation shows that the eigenvalues (and thus the
difference between $\Lambda_0$ and $\Lambda_1$)
decrease as $\sigma_1^2$ when $\sigma_1^2$  is small.  The 
transition time therefore decreases with noise.

\begin{figure}[tbp]
{\hspace*{0.15cm}\psfig{figure=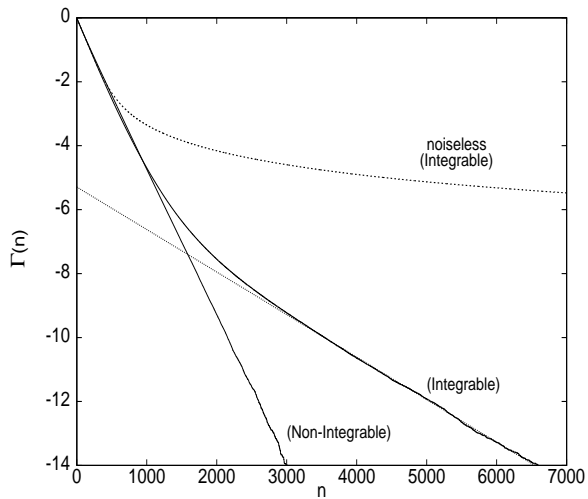,height=6.5cm,width=7.65cm,angle=270}}
{\vspace*{.13in}}
\caption[ty]{
$\Gamma(n)$ (log scale) for the $(\pi/2,\pi/3,\pi/6)$ 
(integrable) and the $(18\pi/31,17\pi/97)$ (non-integrable) triangle
along with the best fit exponential (dashed line) in the
integrable case. Here $\sigma = 0.00005$
and $\Delta q_0 = 0.005$.}

\label{fig:3}
\end{figure}

\noindent
For $\sigma = 0.00005$ (see fig.~\ref{fig:3}), the transition time 
decreases significantly and exponential
decay sets in for $n > 3500$ in the integrable case
while the leading eigenvalue dominates from the
beginning in the non-integrable case. Thus the gap between
$\Lambda_0$ and $\Lambda_1$ increases with noise.

In conclusion, the broad picture that emerges from these
numerical experiments is as follows :

(i) Additive noise makes the spectrum of the evolution
operator discreet.

(ii) When the dynamics is intermittent or regular and the
noise weak, exponential decays may emerge only asymptotically
due to the presence of closely spaced eigenvalues around the
leading eigenvalue, $\Lambda_0$. These are remnants of the
continuous spectrum that exists in the zero noise case.
The transition phase in such a situation can mimic an
algebraic decay.

There are important fallouts of this conclusion. In experimental 
situations where noise is inevitable, signatures of  
exponential decays are not necessarily indicative of chaotic dynamics. 
For instance, semiclassical theory links Lorentzian line shapes 
observed in experiments on ballistic transport in chaotic microstructures
to the exponential decay in the survival fraction \cite{jalabert99}. 
The present analysis however shows that noisy intermittent dynamics
can also give rise to Lorentzian line shapes and it is interesting to
note that there are instances where observations on 
regular or marginally stable
cavities have been found to be no different from chaotic 
cavities \cite{jalabert99}.

\newcommand{\PR}[1]{{Phys.\ Rep.}\/ {\bf #1}}
\newcommand{\PRL}[1]{{Phys.\ Rev.\ Lett.}\/ {\bf #1}}
\newcommand{\PRA}[1]{{Phys.\ Rev.\ A}\/ {\bf #1}}
\newcommand{\PRB}[1]{{Phys.\ Rev.\ B}\/ {\bf #1}}
\newcommand{\PRD}[1]{{Phys.\ Rev.\ D}\/ {\bf #1}}
\newcommand{\PRE}[1]{{Phys.\ Rev.\ E}\/ {\bf #1}}
\newcommand{\JPA}[1]{{J.\ Phys.\ A}\/ {\bf #1}}
\newcommand{\JPB}[1]{{J.\ Phys.\ B}\/ {\bf #1}}
\newcommand{\JCP}[1]{{J.\ Chem.\ Phys.}\/ {\bf #1}}
\newcommand{\JPC}[1]{{J.\ Phys.\ Chem.}\/ {\bf #1}}
\newcommand{\JMP}[1]{{J.\ Math.\ Phys.}\/ {\bf #1}}
\newcommand{\JSP}[1]{{J.\ Stat.\ Phys.}\/ {\bf #1}}
\newcommand{\AP}[1]{{Ann.\ Phys.}\/ {\bf #1}}
\newcommand{\PLB}[1]{{Phys.\ Lett.\ B}\/ {\bf #1}}
\newcommand{\PLA}[1]{{Phys.\ Lett.\ A}\/ {\bf #1}}
\newcommand{\PD}[1]{{Physica D}\/ {\bf #1}}
\newcommand{\NPB}[1]{{Nucl.\ Phys.\ B}\/ {\bf #1}}
\newcommand{\INCB}[1]{{Il Nuov.\ Cim.\ B}\/ {\bf #1}}
\newcommand{\JETP}[1]{{Sov.\ Phys.\ JETP}\/ {\bf #1}}
\newcommand{\JETPL}[1]{{JETP Lett.\ }\/ {\bf #1}}
\newcommand{\RMS}[1]{{Russ.\ Math.\ Surv.}\/ {\bf #1}}
\newcommand{\USSR}[1]{{Math.\ USSR.\ Sb.}\/ {\bf #1}}
\newcommand{\PST}[1]{{Phys.\ Scripta T}\/ {\bf #1}}
\newcommand{\CM}[1]{{Cont.\ Math.}\/ {\bf #1}}
\newcommand{\JMPA}[1]{{J.\ Math.\ Pure Appl.}\/ {\bf #1}}
\newcommand{\CMP}[1]{{Comm.\ Math.\ Phys.}\/ {\bf #1}}
\newcommand{\PRS}[1]{{Proc.\ R.\ Soc. Lond.\ A}\/ {\bf #1}}

\begin{table}[tbp]
\begin{tabular}{cll}
s &     $\Lambda_0$ &   $\Lambda_{fit}$ \\\hline
0.5 & 0.99470 & 0.99467 \\
0.7 & 0.99508 & 0.99502 \\
0.9 & 0.99553 & 0.99546 \\
\end{tabular}
\vskip 0.1 in
\caption{The leading eigenvalue, $\Lambda_0$ computed by
discreetizing $\Lop (x,y)$ compared to the best fit value
$\Lambda_{fit}$ (see fig.~1) for three different parameter
values.
\label{tab}}
\end{table}


\begin{references}
\bibitem{birkhoff}  In Birkhoff co-ordinates, $q$ is measured along
the boundary while $p = \sin(\theta)$ where $\theta$ is
the angle between the ray and the inward normal at
the boundary point $q$.
\end{references}
\end{document}